# Growth and properties of few-layer graphene prepared by chemical vapor deposition

*By Hye Jin Park,[1] Jannik Meyer,[2] Siegmar Roth,[1] Viera Skákalová[1*]*

[1]Max Planck Instititute for Solid State Research, Heisenbergstraße 1, 70569 Stuttgart (Germany)

[2]Electron Microscopy Group of Materials Science, University of Ulm, Albert Einstein Allee 11, 89069 ULM (Germany)

* Corresponding author. Fax: +49 711 689 1010.   E-mail address: v.skakalova@fkf.mpg.de (V. Skákalová)

**ABSTRACT**

The structure, and electrical, mechanical and optical properties of few-layer graphene (FLG) synthesized by chemical vapor deposition (CVD) on a Ni coated substrate were studied. Atomic resolution transmission electron microscope (TEM) images show highly crystalline single layer parts of the sample changing to multilayer domains where crystal boundaries are connected by chemical bonds. This suggests two different growth mechanisms. CVD and carbon segregation participate in the growth process and are responsible for the different structural formations found. Measurements of the electrical and mechanical properties on the centimeter scale provide evidence of a large scale structural continuity: 1) in the temperature dependence of the electrical conductivity, a non-zero value near 0 K indicates the metallic character of electronic transport; 2) the Young's modulus of a pristine polycarbonate film (1.37 GPa) improves significantly when covered with FLG (1.85 GPa). The latter indicates an extraordinary Young modulus value of the FLG-coating of TPa orders of magnitude. Raman and optical spectroscopy support the previous conclusions. The sample can be used as a flexible and transparent electrode and is suitable for special membranes to detect and study individual molecules in high resolution TEM.

# 1. Introduction

The preparation of two-dimensional single-layer atomic crystals [1, 2] and the study of their peculiar properties have lately become a hot issue now that single-layer graphene can be obtained by micro-mechanical cleavage methods [3], showing fascinating physical properties, such as thermodynamic stability, extremely high charge-carrier mobility and mechanical stiffness [4]. Furthermore, graphene has already been showing promises for applications such as transistors [5], transparent electrodes [6-8], liquid crystal devices [9], ultracapacitors [10], and ultra-tough paper [11]. Therefore, the fabrication of graphene sheets of both large area and high quality has become an important challenge. There are several ways to produce single or few-layer graphene, such as mechanical cleavage [3], epitaxy growth in ultrahigh vacuum [12], atmospheric pressure graphitization of SiC [13], chemical oxidation of graphite [14, 15], and chemical vapor deposition (CVD) using transition metals as catalysts [8, 16-21]. Among them, the latter (CVD growth) is considered as one of the most promising, thanks to its flexibility; it could produce not only large-area graphene sheets, but also various chemically tuned 2-dimensional (2D) materials, such as substitutionally doped graphene [22], $^{13}$C-graphene [23], and hexagonal boron nitride films [24], simply by changing gas sources. Growth of graphite using CVD at ambient pressure has been tried for a long time, but the fact that single-layer graphene could be synthesized was proven only very recently by A. Reina et al [17] and K. S. Kim et al [8], independently. Even though they have made a breakthrough by showing that large area single- and few-layer graphene can be synthesized, the growth mechanism and many aspects of the properties of CVD-grown graphene still remain unknown. In this light, we have studied the growth mechanism of graphene sheets using the CVD method and characterized their structure, electrical, optical, and mechanical properties in order to fabricate a high-quality large-area graphene sheet.

2. Experimental section

*2.1*. Few-layer graphene sheet synthesis

Few-layer graphene (FLG) was synthesized by the CVD method, as previously reported [8, 17], with some modifications. The schematic diagram of CVD set-up is provided in Supplementary Information. A 300 nm thick Ni film was coated onto a $SiO_2$/Si substrate using an electron-beam evaporator. The Ni-coated substrate was positioned at the center of a quartz tube and heated to 1000 $^oC$ at a 40 $^oC$/min heating rate, under a flow of argon and hydrogen (Ar/$H_2$=1, 1000 sccm). The substrate remained at 1000 $^oC$ in a Ar/$H_2$ flow for 20 min to anneal the Ni film. The CVD growth of FLG was conducted at 960 ~ 970 $^oC$, using a mixture of gases with a composition ($CH_4$ : Ar : $H_2$ = 250 : 1000 : 4000 sccm). The reaction time was varied from 30 s to 7 min. After the CVD reaction, the sample in the quartz tube was cooled down to 400 $^oC$ at the rate 8.5 $^oC$/min, under a flow of argon and hydrogen. Then the gas valve for hydrogen was closed and the sample was cooled to room temperature under argon atmosphere.

*2.2*. Few-layer graphene/polycarbonate film preparation

Poly(bisphenol A carbonate) (PC) was dissolved in chloroform (solid content: ~15 wt%). The PC/chloroform solution was spin-coated onto the FLG/Ni/$SiO_2$/Si substrate with 500 rpm for 2 min. The coating was homogeneous with ~10 μm thickness. The FLG/PC film was released from the substrate by chemical etching of the Ni layer with a FeCl$_3$ (1 M) solution, followed, in some cases, by a concentrated HCl solution. The etching time varied depending on the substrate size from 30 min to 12 h. The FLG/PC film was rinsed several times with DI-water and dried with nitrogen gas.

*2.3.* Characterization

The FLG was transferred onto arbitrary substrates such as a silicon wafer, a quartz slide, a KBr pellet, and a TEM grid, by selective dissolution of the PC layer from the FLG/PC film, using chloroform. The substrate was chosen depending on the purpose of the characterization. Raman spectra of samples were obtained with a Raman microscope equipped with a Jobin-Yvon LabRam spectrometer with a spectral resolution of 4 cm$^{-1}$. The laser excitation wavelength was 633 nm. All measurements were performed using a 100× objective lens and a D0.3 filter. Optical absorption of the FLG was studied using a UV/Vis/NIR Perkin-Elmer Lambda spectrometer. Scanning electron microscopy (SEM) analysis was carried out with a high resolution FEI XL30 SFEG analytical SEM operated at 0.8 kV or 5 kV. Transmission electron microscopy (TEM) was performed using an imaging-side spherical aberration corrected Titan 80-300 (FEI, Netherlands), operated at 80 kV. Annular dark field scanning transmission electron microscopy (ADF-STEM) was done to determine the number of layers of the FLG on a Quantifoil grid having a 3.5 μm hole. Atomically resolved high-resolution (HR-TEM) images were taken to characterize the atomic structure of the FLG. A value of Cs = 15 μm with an underfocus of ca. 10 nm was used. Atoms appear black at these conditions. Optical microscopy was used to evaluate the layer distribution of an FLG sheet with an area of 4 x 4 mm$^2$ on a silicon wafer having a 300 nm thick thermal oxide layer at the top. The FLG/SiO$_2$/Si samples were also used to study the temperature dependence of electrical conductivity by the 4-probe measurement method. For this purpose, four gold contacts were evaporated on top of the graphene layer. The outer electrodes were connected to a Keithley 2400 source-meter providing a constant current of 1 μA, and the voltage drop between the inner electrodes was recorded with a Keithley 2000 voltmeter. The channel length for the voltage drop was 400 μm. The sample was placed into the sample chamber and annealed at 120 °C in vacuum (10$^{-7}$ mbar) for 24 h before the measurement. The sheet conductance of the sample was measured as a function of temperature from 300 K to 4 K.

Besides the T-dependence measured on the silicon substrate, the sheet resistance of a large-area (2 x 2 cm$^2$) transparent FLG/PC film was measured also in a four-probe configuration, using the Keithley 225 current source and sensitive Keithley 181 digital nanovoltmeters for voltage reading. The Young's modulus of the FLG was determined from the elastic part of the stress vs strain curve, measured by a force transducer Hottinger Baldwin Messtechnik, type 52.

**3. Results and discussion**

We synthesized a continuous FLG sheet on a nickel-coated silicon substrate several square centimetres large. The growth conditions were set starting first with the values reported in references [8, 17-20] and then the growth parameters such as reaction time and ratio of gas mixture between methane and hydrogen, were optimized in order to obtain a continuous and homogeneous coverage of the large-surface substrate. In our case, the optimum volume ratio of methane to hydrogen at atmospheric pressure was empirically set to less than 1/15. It is well-known that hydrogen gas plays a key role for CVD diamond growth, while it acts as an etchant for amorphous carbon [25]. For CVD graphene growth we found that a critical amount of hydrogen is also necessary to synthesize FLG because hydrogen keeps a balance between the production of reactive hydrocarbonaceous radicals and the etching of the graphite layer during the CVD process. If the ratio of methane to hydrogen is too low, the etching reaction becomes much faster than the formation of graphene layers. This was also experimentally proved in recent work of Kong's group [20]. The reaction time was varied between 30 s and 7 min. The SEM image in Fig. 1(a) shows that multi-layer graphene is formed already at a very short reaction time (30 sec). When the electric field between the Ni-substrate and the detector is applied vertically to the sample surface, this will result in a difference in electrical potential along the sample surface, varying according to the number of graphene layers. This is transmitted as a different contrast in the SEM image.

The brightest part in the image corresponds to single-layer graphene having strong coupling to the nickel substrate, while the dark area shows a multi-layers graphene sheet whose projected top layer is electrically less coupled to the nickel layer due to significantly lower electrical conductivity in the z-axis of graphitic crystal. The contrast of the SEM image is reduced as the reaction time increases, since continuous, electrically conductive layers of graphene cover the Ni substrate, so that the potential along the sample surface is more uniform, however, a different number of graphitic layers can still be distinguished (Fig. 1(b)). Raman spectra of the FLG attached to the nickel substrate after synthesis reveal that the number of graphene layers varies from a single layer to several layers, depending on the position of the graphene sheet (Fig. 1(c)). However, the line shape of the D* mode (2600 ~ 2800 cm$^{-1}$) does not change as the number of layers increases, unlike in Raman spectra of samples prepared by the micro-mechanical cleavage method [26]. The full width at half the maximum of the D* peak becomes broader and the peak's position is blue-shifted by up to 20 cm$^{-1}$ as the number of layer increases. The D* mode of a single-layer shows single Lorentzian line-shapes with a line-width in the range 35 ~ 40 cm$^{-1}$.

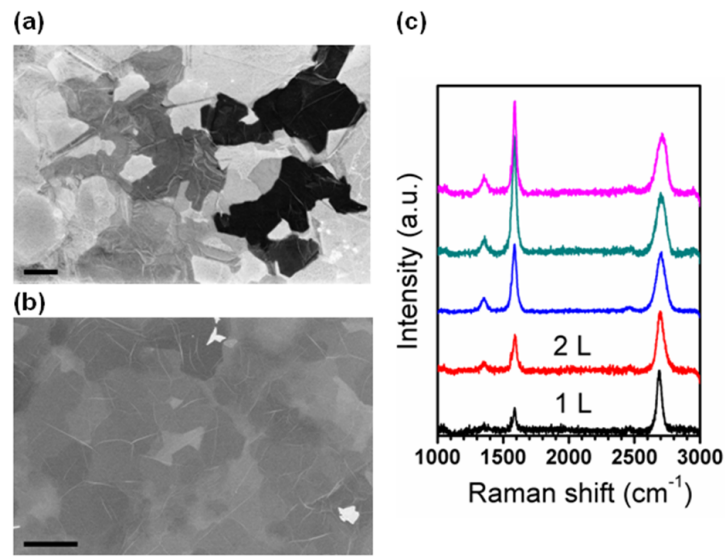

**Fig. 1 - SEM image of a CVD-grown graphene sheet on a nickel-coated SiO$_2$/Si substrate (a) for 30 sec and (b) for 7 min, (c) Raman spectra of a CVD-grown graphene sheet on a nickel coated SiO$_2$/Si obtained from different positions of the sample (scale bar : 1 μm).**

In order to study various physical properties, the FLG samples were then transferred onto various arbitrary substrates such as silicon wafers, quartz glasses and KBr pellets, depending on the characterization method. A FLG was transferred first onto a PC film

(following the procedure described above) and then, either the FLG/PC film itself was characterized, or it served as a medium for the next transferring process. The presence of the FLG on PC film was confirmed by Raman spectra as well (SI-2). The transfer onto the PC film gives the possibility to study the morphology of the back side of the FLG sheet, originally attached to the Ni-layer. The SEM image in Fig. 2(a) clearly shows a large amount of amorphous carbon covering the bottom side of the FLG after the nickel layer was etched by a 1 M $FeCl_3$ solution. The presence of amorphous carbon at the bottom side of the FLG strongly supports a growth mechanism of graphite on nickel by CVD as suggested by Obraztsov et al [16, 27]. On the contrary, the next SEM image (Fig. 2(b)), taken after further treatment using concentrated HCl for 1 h, shows a clean surface of the FLG, demonstrating how efficiently amorphous carbon can be removed by this procedure. The cleaning step developed and demonstrated in this work might be very important in applications for transparent electronic devices.

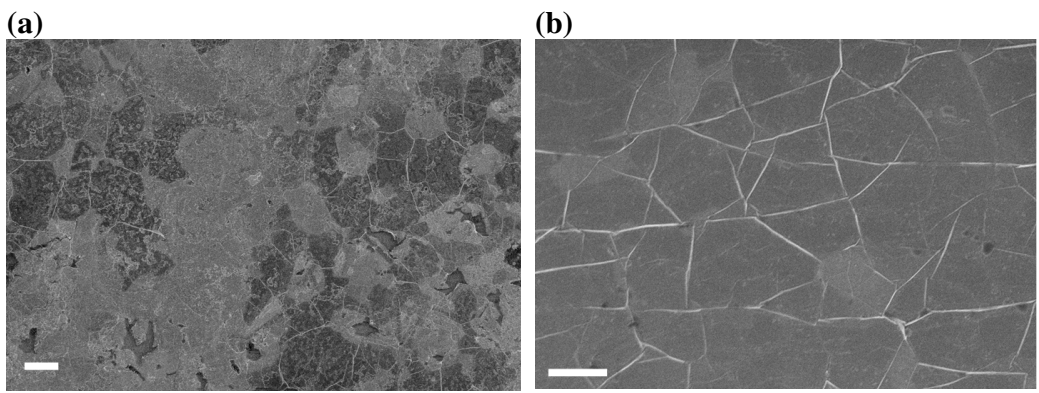

**Fig. 2 -  SEM images of transferred few-layer graphene on polycarbonate film after Ni etching (a) by $FeCl_3$ and (b) by $FeCl_3$ and HCl  (scale bar : 1 μm).**

 In order to study the structure of FLG, we transferred the FLG onto a Quantifoil grid having 3.5 μm holes, by dissolution of the PC layer from the FLG surface, using chloroform. No residue of PC was detected on the sample. Fig. 3 represents an ADF-STEM image of an FLG sample. Determined from electron diffraction, the single layer and bilayer were first identified and their ADF intensities were taken then as a reference for the ADF-STEM thickness analysis using the formula $I \approx t \times Z^{1.7}$, where I is intensity, t is thickness and Z is the atomic number of scattering atoms. Surprisingly, the distribution of the number of graphene layers did not depend much on the duration of the CVD process, so that no significant difference in distribution of the number of layers between the samples synthesized for 30 sec, 1 min, 2 min and 3 min was identified.

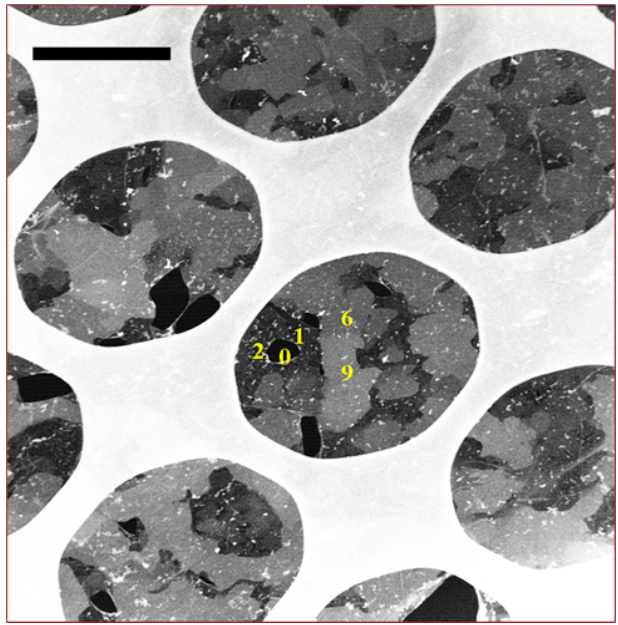

**Fig. 3 - ADF-STEM image of a few-layer graphene sheet on Quantifoil with 3.5 μm hole. The reaction time of the few-layer graphene was 30 seconds. The yellow figures indicate the number of graphene layers (scale bar : 2 μm).**

Each of the samples contained single-layer graphene covering about 2% of the area and few-layer graphene (less than 10 layers) on most of the area. The TEM image in Fig. 4(a) shows a single-layer region of FLG surrounded by a multi-layer region. In some places at the multilayer boundaries, a crystal face of the graphene layer can be distinguished from the angles of the edges. The atomic resolution image of single-layer graphene (Fig. 4(b)) shows a perfect single crystal structure in the displayed region, where no defect can be seen. Fig. 4(c) presents an HR-TEM image of a bilayer region where a clearly pronounced Moiré pattern indicates a turbostratic arrangement of the layers. Overall, we observe a mixture of AB stacked and turbostratically arranged layers. The turbostratic arrangement might happen due to a non-equilibrium process during the FLG growth where two parallel mechanisms, a pure CVD process as well as carbon segregation during fast cooling (which can be referred to as quenching), take place [28]. Under these highly non-equilibrium conditions of fast FLG growth, formation of energetically more favourable tightly stacked graphitic layers is suppressed and, instead, a less ordered turbostratic structure is formed.

Interestingly, in few-layer regions, we found polycrystalline structures with different orientations, but the different domains seemed to be connected by chemical bonds (Fig. 4(d)). This might be responsible for the huge mechanical strength reported below, even though the inhomogeneous structure of the FLG dominates over a large area. Analysis of number of graphitic layers from electron diffraction patterns was discussed in detail in the literature [34,35].

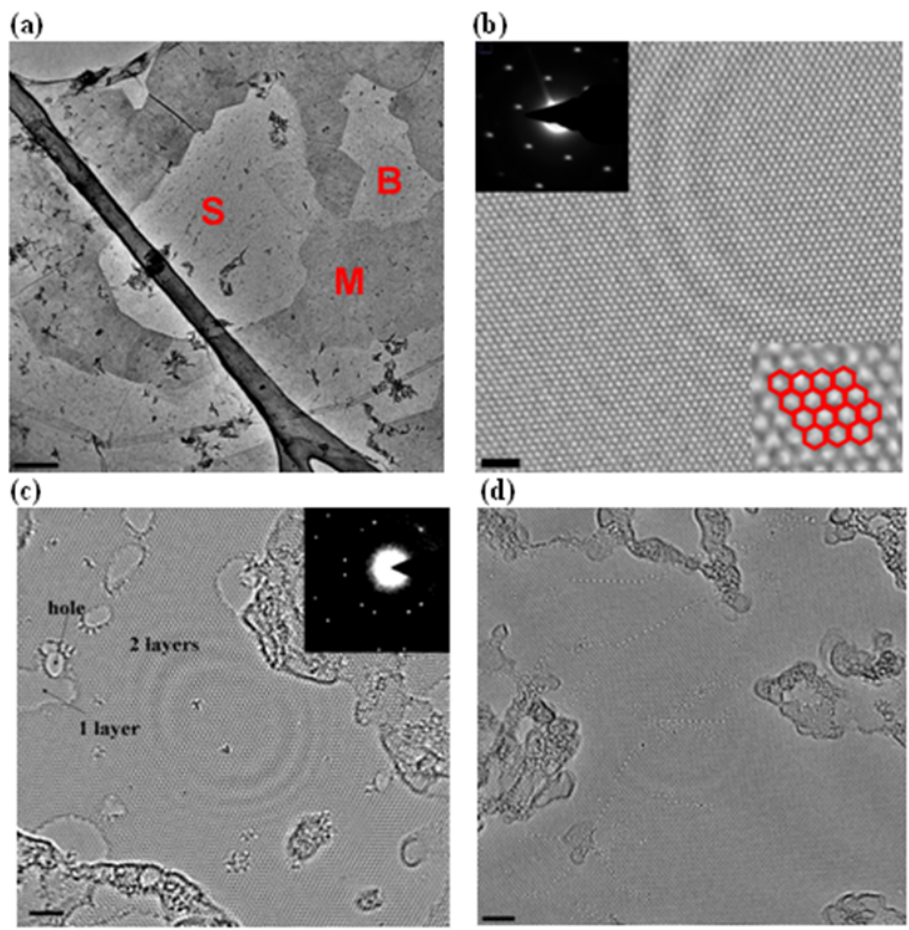

**Fig. 4 - TEM images of few-layer graphene (a) single layer graphene (S) and bilayer graphene (B) supported by multi-layer graphene (M) (scale bar : 250 nm), (b) Atomic resolution image of the single layer in (a) (scale bar : 1 nm), inset: electron diffraction pattern corresponding to a single layer graphene (left) and higher magnification image (right) (c) bilayer graphene having a mismatch of crystal orientation (scale bar : 2 nm), inset : electron diffraction pattern corresponding to a turbostratically arranged bilayer (right) and (d) chemically bonded few-layer graphene having different orientations (scale bar : 2 nm).**

The FLG was also transferred on a quartz slide and on a KBr pellet in order to characterize its optical absorbance behaviour in the IR-NIR-VIS range. In the case of graphene prepared by micro-mechanical cleavage, photon absorbance does not change when the wavelength varies from the visible to the infrared region. The value of the absorbance intensity is given by the fine structure constant or its multiples as the number of graphene layers increases [29, 30]. However, our FLG shows a monotonic increase of absorbance as the photon energy increases up to 4.5 eV where an absorbance maximum appears (Fig. 5(b)). We interpret the increase of absorbance in the FLG as a result of light absorption by the multilayer graphitic domains. Despite the presence of a certain proportion of single-layers in FLG, the character of the optical absorption in the NIR-VIS region is similar with that of a bulk graphite [37].

We measured the sheet resistance of FLG sheets as a function of the temperature R(T). All samples prepared at 3 min of reaction time show almost the same "semiconducting" behaviour: R(T) decreases as the temperature increases (Fig. 5(c)). However, the total change in resistance from 4 K up to room temperature is only a factor of two [31]. For a semiconductor, the energy gap in the electron energy spectrum would cause electrical conductivity to drop to zero as thermal activation is becoming inefficient. In our case the trend of conductance at low temperatures contradicts the semiconductor behaviour: Sheet conductance plotted in the log-log scale (Fig. 5(c): inset) clearly shows convergence of conductance to a non-zero value as T→0 K.

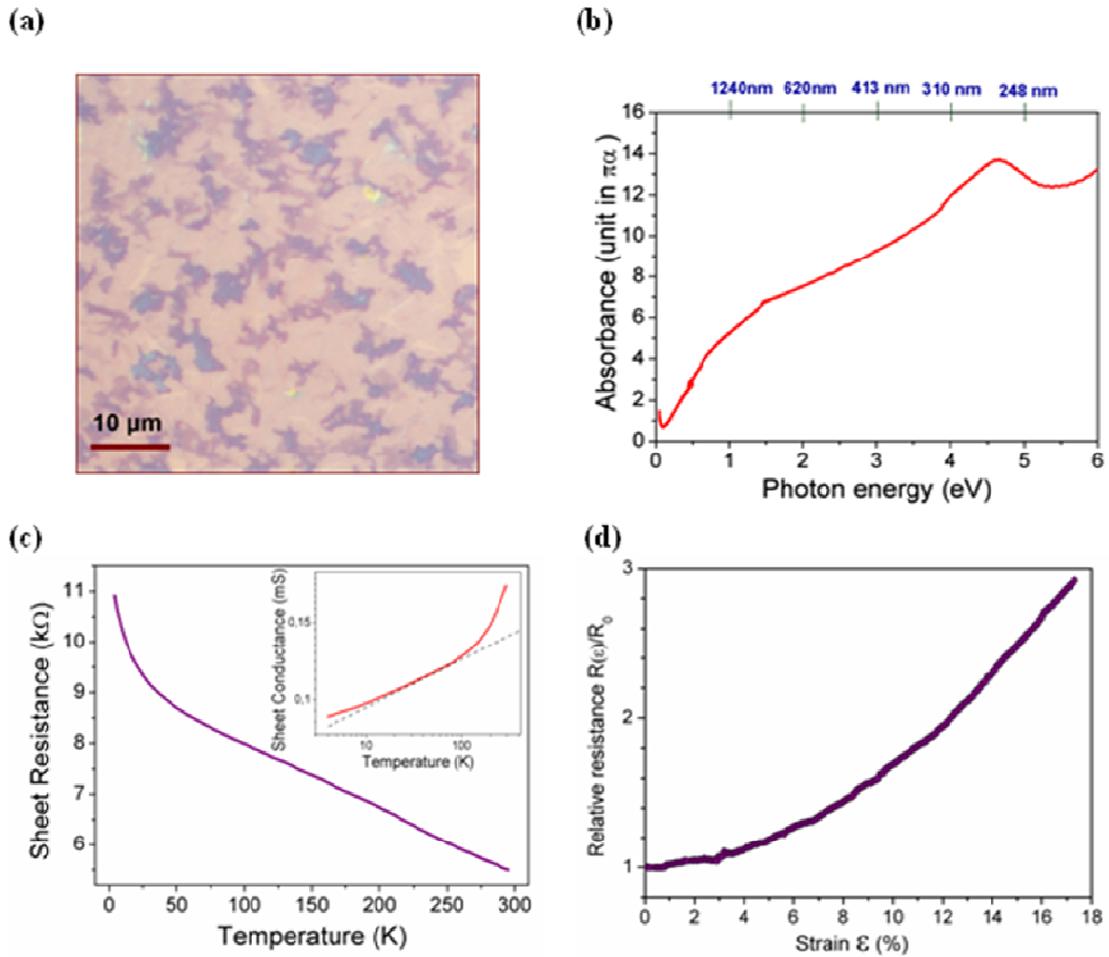

**Fig. 5 - (a) Optical microscopy image of few-layer graphene on SiO$_2$/Si with 300 nm oxide layer (b) optical absorbance of CVD grown graphene depending on photon energy (c) sheet resistance as a function of temperature (inset) sheet conductance as a log-log scale (d) resistance variation of few-layer graphene/PC depending on strain.**

The mechanical properties of the FLG were investigated comparing Young's modulus of a simple PC film (1.37 GPa) with Young's modulus of a double-layer composite FLG/PC film (1.85 GPa). The values were determined from the elastic part of the stress vs strain curve (SI-4(a)). We calculated Young's modulus of the FLG by using a formula for Young's modulus of composed layers [32], considering that our FLG sample consists of five graphene layers (ADF-STEM estimate). Then the value of Young's modulus of the FLG is 4.6 TPa. Based on the HR-TEM investigation, even though the layers of FLG contain many polycrystalline domains, the high value of Young's modulus can be explained by the fact that the different crystal domains merge, thus forming mainly chemical bonds between each other (Fig. 4(d)). Fig. 5(d) shows the electrical sheet resistance, measured also as a function of strain. In the elastic part below the 2% prolongation, there is no observable change in sheet resistance. At higher strain, sheet resistance increases so that it triples its original value when the strain reaches 17%.

Finally, even though there is still a way to go to optimize CVD growth in order to obtain a more homogeneous structure and to eliminate the multilayer "scales" decorating the surfaces of FLG sheets, we can already use the FLG in two applications. Polycarbonate, which is well-known for its high optical transparency and its good mechanical performance, was chosen as a substrate to produce flexible transparent electrodes based on a FLG/PC film. The sheet resistance vs. the transmittance of the films were shown in SI-5. A sample with 84% transmittance measured at a wavelength of 550 nm has 800 $\Omega$/sq of sheet resistance. This value is not as high as that of a doped single-walled carbon nanotubes transparent network produced by the arc-discharge method [33]. However, we believe there is still space for improving the quality of CVD-grown FLG films. The main obstacle limiting the quality of our flexible transparent electrode is the appearance of thick multi-layer "scales" which absorb the visible light (Fig. 5(b)), but do not contribute to the electrical conductivity, since they form small, non-connected islands. The second application we have realized is a special membrane for TEM study. The unique potential of using graphene as an ideal support film for electron microscopy has been demonstrated recently [34]. However, the commonly used method for preparing a graphene membrane, namely through micro-mechanical cleavage, requires the use of a complex process that yields one micron-sized graphene at a time. On the contrary, CVD- grown FLG can be transferred by a single operation on a large number of TEM grids. As we have shown in Fig. 4(b), there are large enough areas of single-layer graphene that can be used for observation, potentially from nano-scale small particles down to single molecules. We show an example in the supporting material (SI-6).

To conclude, we have synthesized an FLG sheet with an area of several square centimetres by the CVD process, and demonstrated its novel properties. It consists of single- and double-layers with randomly distributed isolated multi-layer "scales" which usually are formed near the grain boundaries of polycrystalline nickel during the growth process. Based on HR-TEM we did show that different crystal domains of the FLG are connected by chemical bonds. This, apparently, leads to extraordinary mechanical properties and reasonably small sheet resistance measured at large area (2 x 2 $cm^2$) of our sample despite the macroscopically inhomogeneous structure. The FLG can be used in applications such as a flexible transparent electrode and a special membrane for TEM study.

**Acknowledgements**

We acknowledge support from the Project SALVE of the Deutsche Forschungsgemeinschaft. V.S. acknowledges the contribution of the Center of Excellence CENAMOST (Slovak Research and Development Agency Contract No. VVCE-0049–07) through its support of project APVV-0628-06. We also thank Stephan Schmid and other members of Technology Department at MPI for their important help at substrate preparation, and Thomas Reindl for SEM measurement. We are grateful to Prof. K. von Klitzing and Prof. K. Kern for general supports and to Prof. A. Geim for initiation of this work.

**Appendix A. Supplementary data**

Supplementary data associated with this article can be found, in the online version, at

**REFERENCES**

[1] Sakamoto J, Heijst J, Lukin O, Schlueter AD. Two-dimensional polymers: Just a dream of synthetic chemists? Angew Chem Int Ed. 2009;48(6):1030-69.

[2] Park S, Ruoff RS. Chemical methods for the production of graphenes. Nat Nano. 2009;4(4):217-24.

[3] Novoselov KS, Geim AK, Morozov SV, Jiang D, Zhang Y, Dubonos SV, et al. Electric field effect in atomically thin carbon films. Science. 2004;306(5696):666-9.

[4] Geim AK, Novoselov KS. The rise of graphene. Nat Mater. 2007;6(3):183-91.

[5] Echtermeyer TJ, Lemme MC, Baus M, Szafranek BN, Geim AK, Kurz H. Nonvolatile switching in graphene field-effect devices. IEEE Electr Device Lett. 2008;29(8):952 - 4.

[6] Eda G, Fanchini G, Chhowalla M. Large-area ultrathin films of reduced graphene oxide as a transparent and flexible electronic material. Nat Nano. 2008;3(5):270-4.

[7] Wang X, Zhi L, Mullen K. Transparent, conductive graphene electrodes for dye-sensitized solar cells. Nano Lett. 2008;8(1):323-7.

[8]     Kim KS, Zhao Y, Jang H, Lee SY, Kim JM, Kim KS, et al. Large-scale pattern growth of graphene films for stretchable transparent electrodes. Nature. 2009;457(7230):706-10.

[9]     Blake P, Brimicombe PD, Nair RR, Booth TJ, Jiang D, Schedin F, et al. Graphene-based liquid crystal device. Nano Lett. 2008;8(6):1704-8.

[10]    Stoller MD, Park S, Zhu Y, An J, Ruoff RS. Graphene-based ultracapacitors. Nano Lett. 2008;8(10):3498-502.

[11]    Dikin DA, Stankovich S, Zimney EJ, Piner RD, Dommett GHB, Evmenenko G, et al. Preparation and characterization of graphene oxide paper. Nature. 2007;448(7152):457-60.

[12]    Berger C, Song Z, Li T, Li X, Ogbazghi AY, Feng R, et al. Ultrathin epitaxial graphite: 2D electron gas properties and a route toward graphene-based nanoelectronics. J Phys Chem B. 2004;108(52):19912-6.

[13]    Emtsev KV, Bostwick A, Horn K, Jobst J, Kellogg GL, Ley L, et al. Towards wafer-size graphene layers by atmospheric pressure graphitization of silicon carbide. Nat Mater. 2009;8(3):203-7.

[14]    Stankovich S, Dikin DA, Piner RD, Kohlhaas KA, Kleinhammes A, Jia Y, et al. Synthesis of graphene-based nanosheets via chemical reduction of exfoliated graphite oxide. Carbon. 2007;45(7):1558-65.

[15]    Tung VC, Allen MJ, Yang Y, Kaner RB. High-throughput solution processing of large-scale graphene. Nat Nano. 2009;4(1):25-9.

[16]    Obraztsov AN, Obraztsova EA, Tyurnina AV, Zolotukhin AA. Chemical vapor deposition of thin graphite films of nanometer thickness. Carbon. 2007;45:2017-21.

[17] Reina A, Jia X, Ho J, Nezich D, Son H, Bulovic V, et al. Large area, few-Layer graphene films on arbitrary substrates by chemical vapor deposition. Nano Lett. 2009;9(1):30-5.

[18] Chae SJ, Güneş F, Kim KK, Kim ES, Han GH, Kim SM,et al. Synthesis of large-area graphene layers on poly-nickel substrate by chemical vapor deposition: wrinkle formation. Adv Mater. 2009;21(22):2328-33.

[19] Li X, Cai W, An J, Kim S, Nah J, Yang D, et al. Large-area synthesis of high-quality and uniform graphene films on copper foils. Science. 2009:1171245.

[20] Reina A, Thiele S, Jia X, Bhaviripudi S, Dresselhaus MS, Schaefer JA, et al. Growth of large-area single- and bi-layer graphene by controlled carbon precipitation on polycrystalline Ni surfaces. Nano Res. 2009;2:509-16.

[21] Vázquez de Parga AL, Calleja F, Borca B, Passeggi MCG, Hinarejos JJ, Guinea F, et al. Periodically Rippled Graphene: Growth and Spatially Resolved Electronic Structure. Phys Rev Lett. 2008;100:056807.

[22] Wei D, Liu Y, Wang Y, Zhang H, Huang L, Yu G. Synthesis of N-doped graphene by chemical vapor deposition and its electrical properties. Nano Lett. 2009;9(5):1752-8.

[23] Cai W, Piner RD, Stadermann FJ, Park S, Shaibat MA, Ishii Y, et al. Synthesis and solid-state NMR structural characterization of $^{13}$C-labeled graphite oxide. Science. 2008;321(5897):1815-7.

[24] Phani A. Thin films of boron nitride grown by CVD. Bull Mater Sci. 1994;17(3):219-24.

[25] Park J-H, Sudarshan TS. Chemical vapor deposition. 1st ed. Scarborough: ASM International; 2001.

[26] Ferrari AC, Meyer JC, Scardaci V, Casiraghi C, Lazzeri M, Mauri F, et al. Raman spectrum of graphene and graphene layers. Phys Rev Lett. 2006;97:187401 (1-4).

[27] Johansson A-S, Lu J, Carlsson J-O. TEM ivestigation of CVD graphite on nickel. Thin Solid Films. 1994;252:19-25.

[28] Yu Q, Lian J, Siriponglert S, Li H, Chen YP, Pei S-S. Graphene segregated on Ni surfaces and transferred to insulators. Appl Phys Lett. 2008;93:113103.

[29] Nair RR, Blake P, Grigorenko AN, Novoselov KS, Booth TJ, Stauber T, et al. Fine structure constant defines visual transparency of graphene. Science. 2008;320:1308.

[30] Li ZQ, Henriksen EA, Jiang Z, Hao Z, Martin MC, Kim P, et al. Dirac charge dynamics in graphene by infrared spectroscopy. Nat Phys. 2008;4:532-5.

[31] Bolotin KI, Sikes KJ, Hone J, Stormer HL, Kim P. Temperature-dependent transport in suspended graphene. Phys Rev Lett. 2008;101(9):096802.

[32] Date EHF. Elastic moduli of some composite materials. J Phys D: Appl Phys. 1930;3:778-82.

[33] Geng H-Z, Kim KK, So KP, Lee YS, Chang Y, Lee YH. Effect of acid treatment on carbon nanotube-based flexible transparent conducting films. J Am Chem Soc. 2007;129(25):7758-9.

[34] Meyer JC, Girit CO, Crommie MF, Zettl A. Imaging and dynamics of light atoms and molecules on graphene. Nature. 2008;454:319-22.

[35]    Meyer JC, Geim AK, Katsnelson MI, Novoselov KS, Obergfell D, Roth S, et al. On the roughness of single- and bi-layer graphene membranes. Solid State Comm. 2007;143:101-9.

[36]    Jellison GE, Hunn JD, Lee HN. Measurement of optical functions of highly oriented pyrolytic graphite in the visible. Phy Rev B. 2007;76:085125.